\documentstyle[aps,prl,multicol,epsfig]{revtex}
\begin{document}
\title{Multiple Functionality in Nanotube Transistors}
\author{Fran\c{c}ois L\'{e}onard$^{1}$ and J. Tersoff$^{2}$}
\address{$^{1}$Sandia National Laboratories, MS 9161, Livermore, CA 94551,\\
USA. \\
$^{2}$IBM Research Division, T. J. Watson Research Center, P.O. Box 218,\\
Yorktown Heights, NY 10598, USA.}
\date{\today }
\maketitle

\begin{abstract}
Calculations of quantum transport in a carbon nanotube transistor show that
such a device offers unique functionality. It can operate as a {\it ballistic%
} field-effect transistor, with excellent characteristics even when scaled
to 10 nm dimensions. At larger gate voltages, channel inversion leads to
resonant tunneling through an electrostatically defined nanoscale quantum
dot. Thus the transistor becomes a {\it gated} resonant-tunneling device,
with negative differential resistance at a tunable threshold. For the
dimensions considered here, the device operates in the Coulomb blockade
regime, even at room temperature.
\end{abstract}

\begin{multicols}{2}
\narrowtext

In the quest for nanoscale devices, carbon nanotubes (NTs) \cite{dekker}
have emerged as a promising material with unique electronic and mechanical
properties. Theoretical studies of NT devices have focused primarily on
two-terminal devices \cite%
{PN,pinning,esfarjaniAPL,NDR,odintsov,maksimenko,mehrez}. However, practical
device architectures generally require three-terminal devices; and current
technology is built primarily around field-effect transistors (FETs). A
number of groups have fabricated NT-based FETs \cite%
{tans,martel,dai,tsukagoshi}, and have demonstrated promising transistor
functionality. The devices realized experimentally typically have dimensions
in the micron range. However, the real promise of NTs lies in the
possibility of nanoscale devices. At this size scale, new effects can become
important, creating new problems but also new opportunities.

Here we show that a nanoscale NT FET offers unique functionality that goes
beyond any existing device. The device geometry we consider, shown in
Fig.~1, is a straightforward idealization of that already used for NT FETs %
\cite{tans,martel}, and does not require doping \cite{PN,NDR} or structural
modification \cite{hyldgaard} of the nanotube. Our calculations show that,
for a 10 nm channel length, the device operates as an excellent {\it %
ballistic} transistor, with current saturation at high bias. In addition,
the gate voltage can be used to drive the channel into inversion, defining a
nanoscale ``quantum dot''. Tunneling through a state localized in the dot
gives strong negative differential resistance, even at room temperature. The
FET can thus operate as a {\it gated} resonant tunneling device. This
approach lifts one of the major limitations of resonant tunneling diodes ---
the resonance occurs at a source-drain voltage that can be directly
controlled via the gate voltage \cite{grtd}.

As illustrated in Fig.~1, the device consists of a single-wall,
semiconducting carbon NT. It is embedded in metal contacts on either side,
defining the source and drain. Between the source and drain electrodes, an
insulating dielectric surrounds the NT up to a radius of 10 nm. A
cylindrical gate of radius 10 nm wraps the dielectric and serves to control
the device behavior. In our calculations, the NT and the metals are
separated by a van der Waals distance of 0.3 nm. (The insulator has
dielectric constant $\varepsilon =3.9$, as for SiO$_{2}$, and it is also
separated from the tube by 0.3 nm.)

We treat a zigzag NT of index (17,0), which has radius 0.66 nm and band gap
0.55 eV. The qualitative aspects of our results also apply to other
semiconducting NTs. We use a tight-binding Hamiltonian with one $\pi $
orbital per carbon atom and a nearest-neighbor matrix element of 2.5 eV \cite%
{wildoer}. The metal Fermi level is chosen to be 1 eV below the NT midgap.
(For the NT midgap 4.5 eV below the vacuum level \cite{NTwkfn}, this
corresponds to a metal workfunction of 5.5 eV, roughly that of Au and Pt.)

We first examine the conductance of the device at low source-drain voltage,
as the gate voltage is varied. For a given gate voltage, we obtain the
zero-bias conductance from \cite{datta} 
\begin{equation}
G=\frac{4e^{2}}{h}\int P(E)\left[ -\frac{\partial f(E)}{\partial E}\right]
dE~,  \label{conductance}
\end{equation}%
where the energy $E$ is relative to the Fermi level, $P(E)$ is the electron
transmission probability across the device at energy $E$, and $f(E)$ is the
Fermi function. We calculate $P(E)$ using the quantum-transport procedure of
Ref.~\cite{datta}, which requires the electrostatic potential $\Phi (z)$
along the NT. (Our device is much smaller than observed scattering lengths %
\cite{McEuen}, so there is no need to include scattering other than
reflection by the device itself.) The NT device is divided into three
regions: two semi-infinite ``leads'', and a ``device region'', which is 18.3
nm in length here unless otherwise specified. Within the scattering region
we use the full self-consistent potential $\Phi (z)$. The potentials in the
leads are taken as constant, and equal to the potentials at the boundaries
of the scattering region.

To obtain the self-consistent potential, as in Ref.~\cite{NDR} we start from
a charge $\sigma (z)$, and obtain $\Phi (z)$ by solving Poisson's equation
with boundary conditions set by the source, drain, and gate voltages, with
the dielectric. The electronic charge $\sigma (z)$ is re-calculated using $%
\Phi (z)$, and the procedure is iterated to self-consistency. (Thus charge
transfer and energy alignment between NT and metal electrode is included
self-consistently.) To calculate $\sigma (z)$, the ``scattering region'' is
periodically repeated, the diagonal elements of the Hamiltonian are shifted
by $-e\Phi (z)$, the local density of states on each atomic site is obtained
by direct diagonalization of the Hamiltonian, and the charge on each site is
given by integration of the product of the local density of states and the
Fermi function. (Modifications to include Coulomb blockade effects are
discussed below.) The charge associated with a ``ring'' of atoms is
approximated as uniformly distributed over a length 0.07 nm of the NT
cylinder and over a radial thickness of 0.1 nm. The potential is not
sensitive to the details of this approximation \cite{NDR}. Because the NT
does not form covalent bonds with the metal or dielectric, we focus on the
limit of weak metal-NT coupling. Then the matrix elements of the NT
Hamiltonian are unaffected by the metal, but charge transfer between metal
and NT (and the resulting electrostatic potential) must still be included.

Figure 2a shows the calculated conductance as a function of gate voltage $%
V_{G}$. Three distinct regimes are seen. At low $V_{G}$ (region I in
Fig.~2a), the device exhibits high conductance, while at higher $V_{G}$
(region II in Fig.~2a) the conductance drops to practically zero. These
correspond to the ``on'' and ``off'' states of a FET. In addition, there is
a third regime (region III of Fig.~2a) which has no analog in conventional
transistors. Here the conductance rises sharply with $V_{G}$, and then drops
to practically zero again.

This behavior can be understood in terms of the band diagram of the device
in the respective regimes. Figure 2b shows the band diagram at low $V_{G}$.
The Fermi level falls below the valence-band edge of the NT, due to the high
metal workfunction, so there is substantial free charge (as holes) in the
leads. There is essentially no barrier to hole transport through the channel
of the device, explaining the very high conductance. (The highest
conductance possible is 1 on this scale, corresponding to perfect
transmission through the device.)

With increasing $V_{G}$, the bands in the channel are pulled down in energy,
Fig.~2c. This creates a substantial energy barrier for transport of holes
across the depleted channel, turning off the current. The gate voltage
required to turn off the device could be reduced by increasing the channel
length or reducing the gate radius.

With increasing gate voltage $V_{G}$, conduction-band states are
electrostatically pulled down into the band gap. The resulting quantum
confinement defines a ``quantum dot'' with localized states, Fig.~2d. For
sufficiently large $V_{G}$, the lowest energy level of the quantum dot drops
below the asymptotic valence band edge. Electrons (or equivalently, holes)
can then tunnel through the quantum dot, giving a sharp rise in conductance.
At this point the quantum-dot level is a resonance, rather than truly
localized. The conductance is maximal when the localized level reaches the
Fermi level. Once the level drops more than $\sim k_{B}T$ below the Fermi
level, there are exponentially few holes at the energy of the localized
level, and the conductance drops sharply. This resonant tunneling also leads
to strong negative differential resistance, as described below. Moreover, in
principle one could design a transistor with sharper turn-on, by operating
near the resonance voltage \cite{hyldgaard} (i.e.\ at the transition between
regions II and III of Fig.~2, rather than the normal operation at the
transition between regions I and II).

Due to the small size of the quantum dot, Coulomb blockade (CB) effects play
an important role in the resonant-tunneling regime, even at room
temperature. The single-electron charging energy is roughly $U\sim 0.5$ eV,
much larger than either $k_{B}T$ or the width in energy $\Gamma $ of the
resonant state. (Here $k_{B}T\approx 0.025$ eV. The finite resonance width $%
\Gamma $ arises because the localized state is degenerate in energy with the
continuum of states in the NT leads, and we calculate that  $\Gamma \approx $
0.015 eV). Since $U\gg \Gamma $, the device is in the CB regime, and the
resonant level behaves in many ways as if truly localized.

We can include the most important effects of CB in our calculation in a
relatively simple way. Because of the large value of $U$, there is no need
to consider higher levels beyond the first CB resonance --- they contribute
negligible charge over the range of gate voltage of interest here. The first
CB resonance occurs when the empty and singly-occupied states are
degenerate, which is satisfied when the first one-electron level reaches the
Fermi level. A localized electron has no ``self-interaction'', i.e.\ it does
not feel its own electrostatic potential, so we calculate the potential in
the quantum dot region treating the localized levels as empty. Outside of
this region the states are extended, so we use the charge and potential
calculated with the actual occupancies. Because of the strong spatial
separation of the localized and extended states relevant for transport, this
provides a reasonable approximation to a full self-interaction correction.
We emphasize that this procedure is not a full many-body calculation of the
transport. It is intended to give the {\it position} of the resonance. The
actual conductance values are only suggestive. (The width of the peak in
Fig.~2a corresponds to the gate voltage required to move the localized level
by $\sim $$k_{B}T$ in energy, so it is probably fairly accurate.)

The band diagram of Fig.~2d and the conductance of Fig.~2a (solid line) were
calculated using this approximation. To illustrate the importance of the CB
effect, we also calculated the conductance {\it without} any correction for
CB, as if $U\ll \Gamma $. The result is shown with a dotted line in Fig.~2a.
The conductance peak is much sharper when CB is included, and is shifted to
lower gate voltages.

So far we have considered only the conductance at small source-drain
voltage, where the system is close to equilibrium. To provide a more
complete picture of the device characteristics, we calculated the current at
larger voltage from \cite{datta}%
\begin{equation}
I=\frac{4e}{h}\int P(E)\left[ f(E)-f(E+eV_{D})\right] dE~,  \label{current}
\end{equation}%
where $V_{D}$ is the drain voltage (taking the source as ``ground'').

At any finite drain voltage, the system is not in equilibrium, so we adapt
our calculation of the charge $\sigma (z)$ as described previously \cite{NDR}%
. For large resistance (low transmission probability), the left and right
sides of the device are each in approximate internal equilibrium, but with
Fermi levels that differ by the drain voltage. We therefore calculate the
charge using separate Fermi functions for the two regions. (In between, the
Fermi level is deep in the bandgap, so the details of the cross-over from
one side to the other have no effect on the charge or potential.) In the
opposite limit of high transmission probability, we can approximate the
electrons travelling from left to right as obeying the Fermi distribution of
the left lead, while those moving from right to left obey the distribution
of the right lead.

Figure 3 shows the current as a function of both drain and gate voltages.
The device is a very effective transistor --- for gate voltages
corresponding to regions I and II of Fig.~2, the device exhibits high and
low current respectively, with an abrupt transition between the two regimes.
(For this figure, we use the low-transmission and high-transmission models
for $V_{G}>$ 4 V and $V_{G}<$ 3 V, respectively. A smooth interpolation is
used for intermediate gate voltages.)

At higher gate voltages, the resonant-tunneling peak (region III of Fig.~2a)
appears as a ridge in Fig.~3a. At fixed gate voltage, with increasing drain
voltage the current increases at first, because of the significant tunneling
conductance. The current then reaches a maximum and decreases, as the
resonant level moves into the bandgap of the drain, where there are no
states available for tunneling. This behavior is shown in Fig.~3b. The gate
voltage controls the energy of the resonant level, and so controls the
position of the current maximum. Thus the device provides {\it gated}
resonant tunneling and negative differential resistance.

An important characteristic of conventional transistors is that the current
saturates with increasing drain voltage. Figure 4 shows that our NT device
also displays current saturation. In conventional devices, the saturation is
due to ``pinch-off''. The NT device, however, is ballistic, and the current
is ultimately limited only by the number of available carriers in the leads.
The dotted line in Fig.~4 shows the current in the limit of perfect
transmission through the device. The actual current calculated numerically
is within about 20\% of this ideal value. (For these calculations, we
increased the length of the region treated self-consistently, because of
electric field penetration into the contact region.)

Our results suggest that nanotube devices are excellent candidates for
future nanoscale devices. In addition to their suitability as FETs, they can
provide novel functionality such as room-temperature gated resonant
tunneling. We anticipate that future NT device designs will provide further
improvements in performance and additional new functionality.

F.L. acknowledges support from the NSERC of Canada and from the Office of
Basic Energy Sicences, Division of Materials Sciences, U.S. Department of
Energy under contract No. DE-AC04-94AL85000. Discussions with S. Datta are
gratefully acknowledged.

\begin{figure}[h]
\psfig{file=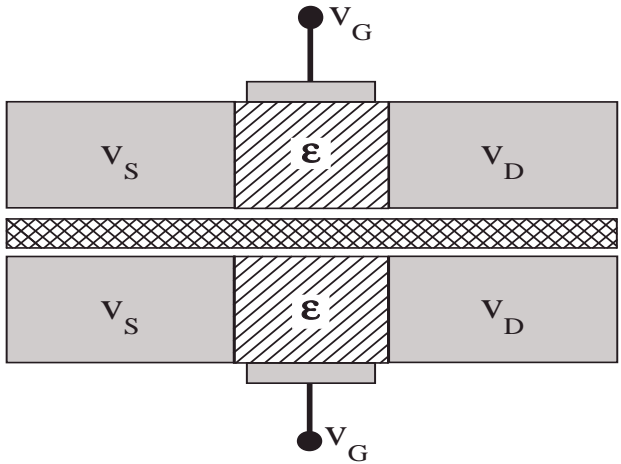,width=240pt,height=180pt}
\caption{Schematic cross-section of the nanotube device. Gray areas are the
gate and the metallic source and drain contacts to the nanotube. Hatched
areas represent the dielectric that surrounds the nanotube, and
cross-hatched area is the nanotube. Source-drain separation is 10 nm;
cylindrical gate has a radius of 10 nm.}
\end{figure}

\begin{figure}[h]
\psfig{file=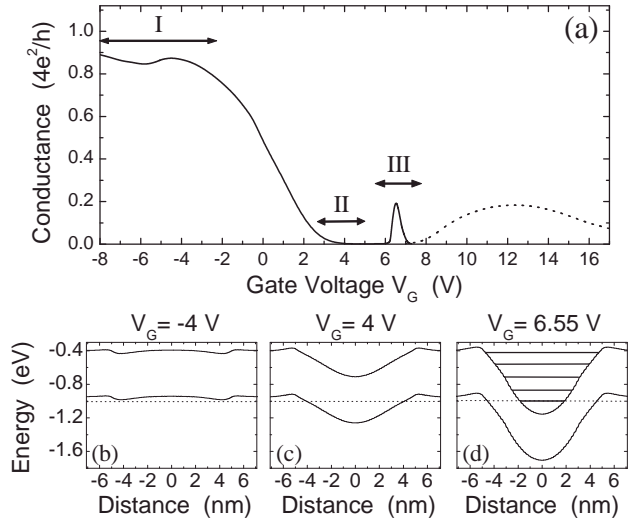,width=240pt,height=200pt}
\caption{(a) Conductance of the nanotube device at low bias. Solid line
includes Coulomb-blockade effects, while dashed line is the result of a
standard self-consistent calculation, as described in text. (b-d) Local
valence and conduction band edges, from the self-consistent electrostatic
potential, for gate voltage indicated. Dotted line is Fermi level. Due to
the high metal workfunction, charge transfer between metal and NT leads to
effective hole doping of the NT in the contacts, with the valence band edge
0.055 eV above the Fermi level. In (c), horizontal lines represent the
single-particle energy levels due to quantum confinement. }
\end{figure}

\begin{figure}[h]
\psfig{file=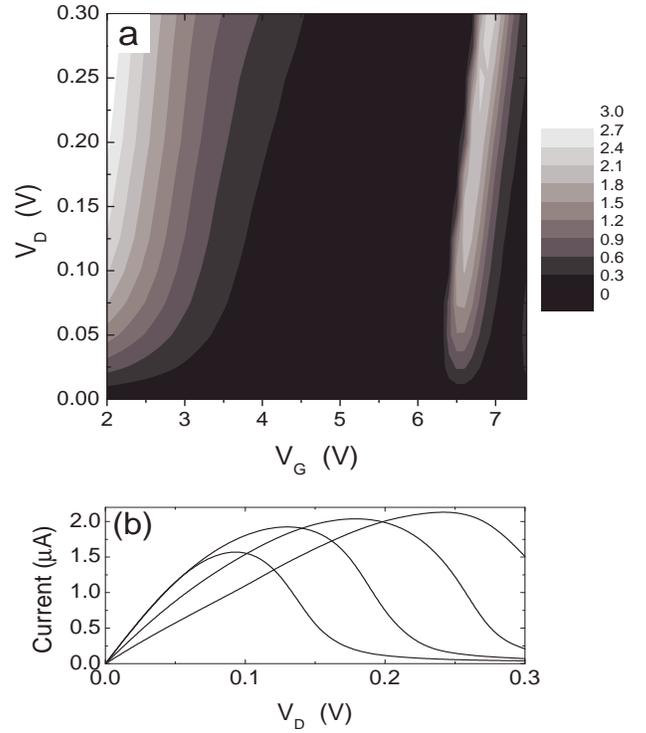,width=240pt,height=280pt}
\caption{(a) Current for the nanotube device (indicated by grey scale, in $%
\protect\mu $A) as a function of drain and gate voltages. (b) Current vs
drain voltage for gate voltages (from left to right) $V_{G}=$6.5, 6.6, 6.7,
and 6.8 V.}
\end{figure}

\begin{figure}[h]
\psfig{file=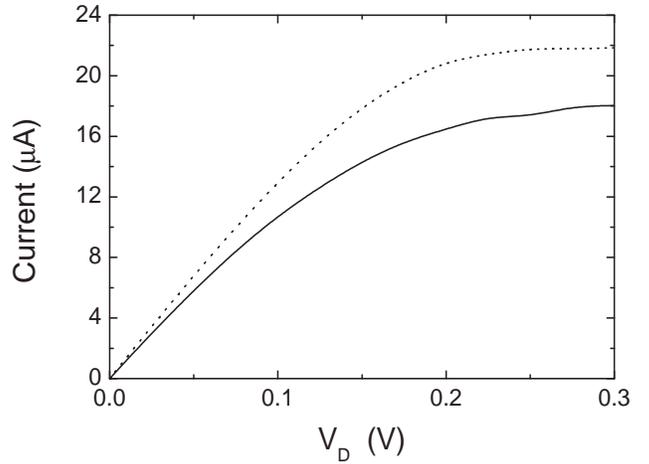,width=240pt,height=180pt}
\caption{Current as a function of drain voltage in the ``on" regime ($%
V_{G}=-2$V). Solid line is numerical result for the high transmission model.
Dotted line is current in the limit of perfect transmission across the
device, for comparison. }
\end{figure}

\end{multicols}

\end{document}